\documentclass[prd,aps,eqsecnum,amsmath,floatfix,nofootinbib,preprint,tightenlines]{revtex4}

\usepackage{latexsym}
\usepackage{graphicx}
\usepackage{multirow}
\usepackage[dvipsnames]{xcolor}
\usepackage{hyperref}

\def\floatcaption#1#2{ \caption{#2 \label{#1}} }

\def\bibi{\bibitem}

\def\a{\alpha}
\def\b{\beta}
\def\c{\chi}
\def\d{\delta}
\def\e{\epsilon}                % Also, \varepsilon
\def\g{\gamma}

\def\l{\lambda}
\def\m{\mu}

\def\p{\pi}                     % Also, \varpi
\def\t{\tau}

\def\D{\Delta}

\def\L{\Lambda}

\def\S{\Sigma}

\def\cl{{\cal L}}
\def\cm{{\cal M}}

\def\cbo{{\,\raise-.15ex\Sc [\,}}                       % curly "
\def\sl#1{\rlap{\hbox{$\mskip 1 mu /$}}#1}      % good slash for lower case
\def\svev#1{\left\langle #1\right\rangle}       % variable < >
\def\ddt#1{{\buildrel {\hbox{\LARGE .\kern-2pt.}} \over {#1}}}% double dot-over
\def\leqx{\,\raisebox{-1.0ex}{$\stackrel{\textstyle <}{\sim}$}\,}

\def\tr{{\rm tr}\,}

\def\ttl#1{{\it #1}}
\def\tb{\tilde\b}
\def\tc{\tilde{c}}

\def\tQ{\tilde{Q}}
\def\tx{\tilde{x}}
\def\tD{\tilde{\Delta}}
\def\tcl{\tilde\cl}

\def\hB{\hat{B}}
\def\hf{\hat{f}}

\def\SU{{\rm SU}}

\begin{document}

\begin{boldmath}
\begin{center}
{\large\bf
The large-mass regime of the dilaton-pion low-energy\\[2 mm] effective theory
}\\[8mm]
Maarten Golterman$^{a,b}$ and Yigal Shamir$^c$\\[8 mm]
{\small
$^a$Department of Physics and Astronomy, San Francisco State University,\\
San Francisco, CA 94132, USA\\
%\\[5mm]
$^b$Department of Physics and IFAE-BIST, Universitat Aut\`onoma de Barcelona\\
E-08193 Bellaterra, Barcelona, Spain\\
$^c$Raymond and Beverly Sackler School of Physics and Astronomy,\\
Tel~Aviv University, 69978, Tel~Aviv, Israel}\\[10mm]
\end{center}
\end{boldmath}

\begin{quotation}
Numerical data of the SU(3) gauge theory with $N_f=8$
fermions in the fundamental representation suggest the existence
of a large-mass regime, where
the fermion mass is not small
relative to the confinement scale,
but nevertheless the dilaton-pion low-energy theory is applicable thanks to the
parametric proximity of the conformal window.  In this regime,
the leading hyperscaling relations are similar to those of
a mass-deformed conformal theory, so that distinguishing infrared conformality
from confinement requires the study of subleading effects.
Assuming that the $N_f=8$ theory confines, we estimate how light
the fermion mass should be to enter the small-mass regime,
where the pions become much lighter than the dilatonic scalar meson.
\end{quotation}

%%%%%%%%%%%%%%%%%%%%%%%%%%%
\newpage
\section{\label{intro} Introduction}
%%%%%%%%%%%%%%%%%%%%%%%%%%%
Asymptotically free gauge theories coupled to a large number of fermions
can have a very small beta function, slowing down the running of the
coupling over a wide energy range, turning it into a ``walking'' coupling.
In the chiral limit, these theories
exist in one of two phases.\footnote{
  Recently a third phase was proposed in Ref.~\cite{CT}.
  For an attempt to apply our approach in the context of QCD,
  and compare it with that of Ref.~\cite{CT}, see Ref.~\cite{nucl}.
}
One option is that the walking coupling
eventually becomes large enough to trigger confinement and
chiral symmetry breaking.  The alternative is that the running
of the coupling comes to a halt, indicating the existence of an
infrared attractive fixed point.  The deep infrared dynamics is then
scale-free, and characterized by power-law correlation functions.

$\SU(N_c)$ gauge theories with $N_f$ Dirac fermions
in the fundamental representation have been the subject of extensive
lattice studies (for $N_c=2,3$), and both types of behavior are expected
to occur, depending on the value of $N_f$.\footnote{%
  For recent reviews, see Refs.~\cite{DeGrandreview,NP,Pica,Benlat17}.
}
The so-called conformal window
then occupies the range $N_f^*(N_c) \le N_f < 11N_c/2$,
where $N_f^*(N_c)$ is defined as the smallest number of flavors
for which the massless $\SU(N_c)$ theory is infrared conformal.

When the running is very slow, it can be extremely challenging
to determine by numerical simulations whether
the massless theory is ultimately confining or infrared conformal.
Nevertheless, evidence is growing that walking theories
just below the conformal window exhibit a light flavor-singlet scalar meson.
In Ref.~\cite{PP} we developed a low-energy effective theory for such theories
which simultaneously
accounts for the usual pions as well as for the flavor-singlet scalar.
Pions are the (pseudo) Nambu-Goldstone bosons arising
from the spontaneous breaking of (approximate) chiral symmetry.
By analogy, we attribute the existence of the light flavor-singlet scalar,
or ``dilatonic meson,'' to the small explicit breaking
of scale invariance by the walking coupling
at the scale where chiral symmetry breaks spontaneously.
This explicit breaking keeps getting smaller as $N_f$ approaches
$N_f^*(N_c)$ from below \cite{PP,latt16}.

In order to turn the proximity of the conformal window into
a continuous parameter of the low-energy theory,
we invoked the Veneziano limit \cite{VZlimit},
where the number of fundamental-representation flavors $N_f$
tends to infinity in proportion to the number of colors $N_c=N$.
One expects this
effective theory to be organized in terms of a systematic power counting in
\begin{equation}
  p^2/\L^2 \ \sim \ m/\L
  \ \sim \ 1/N \ \sim \ |n_f-n_f^*| \ \sim \ \d \ll 1 \ .
\label{pc}
\end{equation}
Here $n_f=N_f/N_c$, and
\begin{equation}
  n_f^* = \lim_{N_c\to\infty} \frac{N_f^*(N_c)}{N_c} \ ,
\label{nfstar}
\end{equation}
is the location of the sill of the conformal window in the Veneziano limit.
As usual, $m$ is the fermion mass, and $p^2$ is a generic external momentum
of order the pion mass squared, while
$\L$ characterizes the confinement scale of the massless theory.

Numerical studies of walking theories reveal a physical behavior which is
qualitatively different from QCD not only in the presence of the
dilatonic meson, but also in other important ways \cite{LSD,LatKMI,sextet}.
First attempts to describe this behavior using the effective theory
suggest, moreover, that the ratio $m/\L$ may not be small in these simulations
\cite{AIP1,AIP2}.  The goal of this paper is to investigate this possibility.
We stress that our investigation is based on the premise
that the theories under
consideration are confining in the infrared; what it will take to confirm
this assumption is a question to which we will return below.

In Sec.~\ref{largeM} we identify a ``large-mass regime,''
and show that, in this regime, the effective theory reproduces
the hyperscaling relations of a mass-deformed conformal theory
to leading order.  We also show that this regime is amenable
to a systematic treatment thanks to the proximity of the conformal sill,
even though the ratio $m/\L$ may not be small.
The small parameter controlling this regime is $n_f-n_f^*$.
In Sec.~\ref{LSD} we revisit the numerical data
of the $\SU(3)$ gauge theory with $N_f=8$ flavors \cite{LSD,LatKMI,AIP1,AIP2}.
Our conclusions are summarized in Sec.~\ref{conc}.
The two appendices are devoted to technical details.

%%%%%%%%%%%%%%%%%%%%%%%%%%%
%\newpage
\section{\label{largeM} The large-mass regime of the effective theory}
%%%%%%%%%%%%%%%%%%%%%%%%%%%
Since the publication of Ref.~\cite{PP}, we came to realize that the
construction of the tree-level lagrangian and of
the associated classical solution are more involved.\footnote{
  We thank R.\ Rattazzi for useful discussions of this issue.
  App.~\ref{classical} corrects a few technical statements made in Sec.~IV of
  Ref.~\cite{PP}.
}
We revisit these steps in App.~\ref{classical}, which also includes a full list
of the technical assumptions needed for the construction.
The end result is the following tree-level lagrangian
\begin{equation}
  \cl = \cl_\p + \cl_\t + \cl_m + \cl_d \ ,
\label{Leff}
\end{equation}
where
\begin{eqnarray}
  \cl_\p &=& \frac{\hf_\p^2}{4}\, e^{2\t}
            \tr(\partial_\m \S^\dagger \partial_\m \S) \ ,
\label{Lp}\\
  \cl_\t &=& \frac{\hf_\t^2}{2}\, e^{2\t} (\partial_\m \t)^2  \ ,
\label{Lt}\\
  \cl_m &=& -\frac{\hf_\p^2 \hB_\p m}{2} \, e^{y\t}
  \tr\Big( \S + \S^\dagger \Big) \ ,
\label{Lm}\\
  \cl_d &=& \hf_\t^2 \hB_\t \, e^{4\t} c_1(\t-1/4) \ ,
\label{Ld}
\end{eqnarray}
and $\g_*=3-y$ is the fixed-point value of the mass anomalous dimension
at the sill of the conformal window \cite{PP,gammay}.
$\cl_\p$ and $\cl_\t$ are kinetic terms for the pions and for the
dilatonic meson, respectively, while $\cl_M$ and $\cl_d$ are
the corresponding potential terms.
The parameter $c_1$ in Eq.~(\ref{Ld}) is proportional to $n_f-n_f^*$.
Here $\S(x)\in \SU(N_f)$ is the usual non-linear field describing the pions,
while $\t(x)$ is a (dimensionless) effective field describing the dilatonic meson.
In order to bring the lagrangian into this form, we shifted the $\t$ field by an amount
$\D_{tot}=\D+\tD$.  Details on the two shifts may be found in App.~\ref{classical}.
The hatted low-energy constants are related to their counterparts
before the $\t$ shifts (see Eq.~(\ref{LeffL})) according to
\begin{equation}
\label{hatLECs}
  \hf_{\p,\t} = e^{\D_{tot}} f_{\p,\t} \ , \quad
  \hB_\p = e^{(y-2)\D_{tot}}B_\p \ , \quad
  \hB_\t = e^{2\D_{tot}}B_\t \ .
\end{equation}
In Eq.~(\ref{Ld}), the presence of the factor $\t-1/4$,
with the specific constant $-1/4$, is a result of the second $\t$ shift,
which was done for convenience.  With this choice, the classical dilatonic-meson vacuum
$v=\svev\t$, which is a function of the fermion mass, $v=v(m)$,
vanishes in the massless limit,
\begin{equation}
\label{v}
  v(0) = 0 \ .
\end{equation}

We now begin our study of the large-mass regime.
For $m>0$, the classical vacuum is the solution of the saddle-point equation
\begin{equation}
\label{LOvev}
\frac{ym}{4c_1\cm}=ve^{(1+\g_*)v}\ ,
\end{equation}
where
\begin{equation}
\label{calM}
\cm=\frac{\hf_\t^2\hB_\t}{\hf_\p^2\hB_\p N_f}\ .
\end{equation}
Before we move on we need to clarify a technical point.
Recall that the mass term of the usual chiral lagrangian involves
the product $B m$, where $B$ is a low-energy constant akin to $\hB_\p$.
Only the product $B m$ has an invariant meaning,
while the determination of $B$ and $m$ separately requires
a renormalization prescription.  Similarly, Eq.~(\ref{Ld})
involves the product $\hB_\t c_1$, where, by analogy with the familiar
chiral case, $\hB_\t$ is a dimensionful low-energy constant
that characterizes the massless theory, while $c_1$
is dimensionless, and proportional to the small parameter $n_f-n_f^*$.
Once again, only the product $\hB_\t c_1$ has a well-defined value
for a given theory,
whereas $\hB_\t$ and $c_1$ cannot be determined separately.

It follows that the ratio $m/(c_1\cm)$,
which occurs on the left-hand side of Eq.~(\ref{LOvev}),
has a well-defined meaning, because it can be expressed
in terms of the products $\hB_\p m$ and $\hB_\t c_1$.
If we write $m/(c_1\cm)=(m/\L)(\L/(c_1\cm))$,
the basic power counting~(\ref{pc}) tells us that
this ratio is of order $\d^0$.  But, because $m$ and $n_f-n_f^*$ are independent
expansion parameters, $m/(c_1\cm)$ can still take small or large values.
In this paper we are interested in the regime where
\begin{equation}
\label{largemass}
  m/(c_1\cm)\gg 1 \ ,
\end{equation}
which we will call the large-mass regime.
As for the individual expansion parameters,
we will assume that always $n_f-n_f^* \ll 1$,
but we will leave the ratio $m/\L$ unspecified,\footnote{
  A possible definition of $\L$ and its determination
  from numerical data are discussed in Sec.~\ref{conc}.
}
thereby exploring whether the effective theory is still controlled by
a systematic expansion when $m/\L$ is not  small.
That the effective theory turns out to be applicable under these circumstances
is somewhat counter-intuitive, since the usual
chiral expansion is valid for small $m/\L$ only.
The key observation, which we discuss in detail below, is
that the loop expansion of the effective theory
is still governed by the small parameter $n_f-n_f^*$ also when $m/\L$ is large.
In fact, as usual, the requirement for a consistent power counting
is really that the loop-expansion parameters, $M_\p^2/(4\p F_\p)^2$
and $M_\t^2/(4\p F_\t)^2$, should be small.
Here $M_\p$ ($M_\t$) is the mass of the pion (dilatonic meson),
while $F_\p$ and $F_\t$ are the corresponding decay constants,
all of which are functions of the input fermion mass $m$.
As we will see, the requirement that the loop-expansion parameters be small
turns out not to be the same as $m/\L$ small.

When the left-hand side of Eq.~(\ref{LOvev}) is large, the dominant dependence
on $v$ on the right-hand side is through the exponential factor.
We may thus approximate Eq.~(\ref{LOvev}) by
\begin{equation}
\label{LOvevx}
\frac{ym}{4c_1\cm} \ \sim \ e^{(1+\g_*)v}\ ,
\end{equation}
hence
\begin{equation}
\label{largemsol}
e^{v(m)} \ \sim \ \left(\frac{ym}{4c_1\cm}\right)^{\frac{1}{1+\g_*}}\ .
\end{equation}
Corrections to this approximate solution for $v(m)$ are of order
$\log v(m) \sim \log \log m$, see App.~\ref{v012}.\footnote{
  Of course, in real fits to numerical data, one would use
  the solution of Eq.~(\ref{LOvev}).
}

Let us now examine how various physical quantities scale with $m$
in this large-mass regime.  We begin with the decay constants.
Much like the relation between the unhatted and hatted low-energy constants
(Eq.~(\ref{hatLECs})), for a given $m$ the physical decay constants are
given at leading order by
\begin{equation}
\label{defF}
  F_{\p,\t} = e^{v(m)} \hf_{\p,\t} \ .
\end{equation}
The masses of non-Nambu-Goldstone hadrons are expected to behave similarly.
For example, the tree-level lagrangian for the nucleon is
\begin{equation}
\label{nucleon}
  \cl_N = \overline{N}({\sl{\partial}} + e^\t m_N) N\ ,
\end{equation}
where $m_N$ is a low-energy constant.
This form follows from the behavior
of the effective nucleon field under a (classical)
scale transformation:  $N(x)\to\l^{3/2}N(\l x)$.  It follows that
the nucleon mass is given by
\begin{equation}
\label{Nmass}
  M_N = e^{v(m)}m_N\ .
\end{equation}
With Eq.~(\ref{largemsol}), we see that the decay constants and the nucleon mass
all satisfy the familiar hyperscaling relation,
\begin{equation}
\label{hyper}
  F_\p, F_\t, M_N \sim m^{\frac{1}{1+\g_*}} \ .
\end{equation}
Turning to the Nambu-Goldstone sector, the pion mass is given by \cite{PP}
\begin{equation}
  M_\p^2 = 2 \hB_\p m e^{(1-\g_*)v} \ ,
\label{mintreepion}
\end{equation}
while the mass of the dilatonic meson is
\begin{equation}
  M_\t^2 = 4c_1\hB_\t e^{2v} (1+(1+\g_*)v) \ .
\label{mintreetau}
\end{equation}
Using Eq.~(\ref{largemsol}) it is easy to see that the Nambu-Goldstone masses
satisfy the same hyperscaling relation as well.

The Nambu-Goldstone boson masses remain parametrically smaller
than decay constants and other masses.
In order to see this, we reintroduce the dependence on $c_1$.  We find
\begin{equation}
\label{MNGB}
  M_\p^2, M_\t^2 \sim c_1^{\frac{-1+\g_*}{1+\g_*}} m^{\frac{2}{1+\g_*}} \ ,
\end{equation}
whereas for the other dimensionful quantities we find
\begin{equation}
\label{MnonNGB}
  F_\p^2, F_\t^2, M_N^2 \sim c_1^{-\frac{2}{1+\g_*}} m^{\frac{2}{1+\g_*}} \ .
\end{equation}
The mass-squared of the Nambu-Goldstone bosons is smaller by a factor of order
\begin{eqnarray}
  c_1^{\frac{-1+\g_*}{1+\g_*}+\frac{2}{1+\g_*}} = c_1 \ ,
\nonumber
\end{eqnarray}
relative to the other dimensionful quantities.
The loop-expansion parameters are therefore parametrically of order $c_1$
as well.

Let us examine the loop-expansion parameters in more detail.
Using the exact saddle-point equation~(\ref{LOvev}) together with
Eq.~(\ref{mintreepion}) gives\footnote{
  There are also loop corrections that involve $M_\p^2/(4\p F_\t)^2$,
  which scales in the same way as $M_\p^2/(4\p F_\p)^2$.
}
\begin{eqnarray}
\label{smallpar}
  \frac{M_\p^2}{(4\p F_\p)^2} &=& \frac{8\hB_\p\cm}{(4\p \hf_\p)^2y}\,c_1 v
\\
  &\sim& \frac{2\hB_\p\cm}{(4\p \hf_\p)^2}\,
  c_1 \log \frac{m}{c_1\cm} \ ,
\nonumber
\end{eqnarray}
where in the last line we used the approximate
large-mass solution~(\ref{largemsol}).
Similarly, for the dilatonic meson we find
\begin{eqnarray}
\label{smallpar2}
  \frac{M_\t^2}{(4\p F_\t)^2}
  &=& \frac{4\hB_\t}{(4\p \hf_\t)^2}\,c_1 (1+(4-y)v)\ ,
\\
  &\sim& \frac{4\hB_\t}{(4\p \hf_\t)^2}\, c_1 \log \frac{m}{c_1\cm} \ ,
\nonumber
\end{eqnarray}
We see that both loop-expansion parameters will be small provided that
\begin{equation}
\label{c1logm}
  c_1 \log \frac{m}{c_1\cm} \ll 1 \ .
\end{equation}
This new constraint sets the range of applicability of the effective theory
in the large-mass regime defined by Eq.~(\ref{largemass}).

The next-to-leading and higher order terms in the lagrangian
of the effective theory serve as counterterms for the loop diagrams
generated using lower-order vertices.  Therefore, these higher-order terms
should follow the same parametric dependence on $m$ and $n_f-n_f^*$
(with the latter represented here by $c_1$)
as the loop-expansion parameters discussed above.  Let us verify that this
is indeed the case.
In ordinary chiral perturbation theory every operator takes the form of
\begin{equation}
  \tQ = \tQ(\S,m,\partial_\m) \ ,
\label{Qtilde}
\end{equation}
where, for simplicity, we have substituted $\chi_{ij} = m \d_{ij}$
for the usual chiral source.  This operator is mapped
into the lagrangian of the dilaton-pion effective theory as \cite{PP}
\begin{equation}
  Q = e^{4\t}\, \tQ(\S,e^{-(1+\g_*)\t}m,e^{-\t}\partial_\m) \ .
\label{Qexp4t}
\end{equation}
It follows that every insertion of $m$ in ordinary chiral perturbation theory
gets replaced in the dilaton-pion effective theory by an insertion of
\begin{equation}
\label{mvm}
  m\, e^{-(1+\g_*)v(m)} \ ,
\end{equation}
where we recall that $v(m)$ is the classical solution.
For $m \leqx c_1\cm$, $v(m)$ is small,
and the modification is innocuous.
The factor of $e^{-(1+\g_*)v}$ is close to one,
and can be re-expanded in a power series in $m$.
In the large-mass regime, on the other hand, $e^{-(1+\g_*)v}$
is much smaller than one.  Indeed, using once again the exact
saddle-point equation~(\ref{LOvev}) gives
\begin{equation}
\label{mvmc1}
  m e^{-(1+\g_*)v} = \frac{4c_1\cm v}{y}
  \sim c_1\cm \log \frac{m}{c_1\cm} \ .
\end{equation}
As expected, the parametric dependence of this expression on $m$ and on $c_1$
is the same as in Eqs.~(\ref{smallpar}) and~(\ref{smallpar2}).

In summary, while for the case of QCD, chiral perturbation theory is an
expansion in powers of $m$, we see that, in the large-mass regime
of the dilaton-pion effective theory, the expansion in powers of $m$
is effectively replaced by an expansion in powers of
$c_1 \log (m/(c_1 \cm))$.  The range of validity of the expansion
is set by condition~(\ref{c1logm}).

Before we continue let us recall that,
if we keep decreasing the fermion mass, we will eventually reach
the small-mass regime, where, by definition, $m/(c_1\cm)\ll 1$.
In this regime the mass of the dilatonic meson freezes out.
The dilatonic meson decouples from the low-energy physics of the pions
(along with all other heavier hadrons),
leaving the pions as the only light degrees of freedom.
The small-mass regime can thus be described by
ordinary chiral perturbation theory as well.\footnote{
  Of course, the dilatonic meson is still much lighter than other hadrons
  in this small-mass regime, and the full effective theory~(\ref{Leff})
  including the dilatonic meson can still be used.
}

Since the dependence of condition~(\ref{c1logm}) on $m$ is only logarithmic,
the large-mass regime can extend over many scales.
It is therefore interesting to
discuss how the expansion parameters renormalize.
The renormalization of $m$ is standard, but that of $c_1$, of course, is not.
As explained in detail in Ref.~\cite{PP}, $c_1$ originates from a single insertion
of the $m$-independent part of the trace anomaly.
As long as $m$ is not large relative to the infrared scale
of the massless theory, $\L$, one has
\begin{equation}
\label{c1btilde}
  c_1 \sim \tb(\L) \sim n_f-n_f^* \ ,
\end{equation}
where $\tb(\m) = (4\a)^{-1}\, \partial\a/\partial\log\m$,
and $\a(\m)=g^2(\m)N_c/(4\p)$ is the renormalized 't~Hooft coupling.
The first approximate equality follows from matching correlation functions
of the microscopic and effective theories, while the second represents
a central dynamical assumption made in Ref.~\cite{PP}.  In itself,
$\a(\L)$ must be large for chiral symmetry breaking to take place.
The smallness of the beta function comes from the proximity
of the conformal window, which is accounted for in the effective theory
by the small parameter $n_f-n_f^*$.
Now, when we consider the effective theory
at a different, and possibly much higher, renormalization scale $\m$,
the first approximately equality remains valid, namely,
\begin{equation}
\label{c1mu}
  c_1(\m) \sim \tb(\m) \ .
\end{equation}
But now we do not necessarily have that $\tb(\m) \sim n_f-n_f^*$.
As the coupling $\a(\m)$ decreases with increasing $\m$,
at first the beta function grows (in absolute value), before it turns around
and approaches zero at the asymptotically free fixed point.   The reason for
this growth is that, for small $m$, the theory is near the infrared attractive
fixed point at a coupling $\a_*$ at $n_f=n_f^*$, and $\a_*$
is just a little larger than
the coupling $\a(\L)$ at which chiral symmetry breaks.\footnote{
  For a heuristic discussion based on the two-loop beta function
  and the gap equation, see Ref.~\cite{latt16}.  The scenario we describe here
  corresponds to the $N_f=12$ theory in Fig.~1 therein, provided that
  the critical coupling were a little smaller, say, $g_c^2=9$
  (instead of $g_c^2=\p^2$), so that according to the model of Ref.~\cite{latt16}
  the $N_f=12$ theory would become walking and confining.
}

This observation suggests that the breakdown of the expansion
in the large-mass regime might alternatively be triggered by the
logarithmic growth of $c_1$ with the renormalization scale $\m$,
because in the large-mass regime we should take $\m\sim m$.
In fact, this does not happen.
If the logarithmic evolution has taken $c_1$ from $c_1(\m=\L) \ll 1$
to $c_1(\m=m) \sim 1$, then, necessarily, $\log m/\cm$ must be large,
which, in turn, implies that $c_1 \log (m/(c_1 \cm))$ is large.
Hence, condition~(\ref{c1logm}) has already been violated at a lower scale,
where $c_1$ is still small.
This implies that the breakdown of the expansion is always triggered by
the failure of condition~(\ref{c1logm}), and that always $c_1\ll 1$ in
the regime of validity of the effective theory.

For completeness, we note that the expansion would also break down
if $m$ grows so much that $\a(\m=m)$ has become too weak to support any
bound states in the first place.  Unfortunately, we are not aware
of any simple criterion that will tell us when this happens from within the
effective theory.

Finally, let us consider the dependence on $N_c$ and $N_f$.
Recall that the decay constants scale as $F_\p \sim \sqrt{N_c}$ and
$F_\t \sim N_c$, while the low-energy constants $\hB_\p$ and $\hB_\t$
are $O(1)$ in the Veneziano limit, from which it follows that
$\cm$ is $O(1)$ in the Veneziano limit as well.
The $N_f$ dependence at one loop was discussed
in Ref.~\cite{AIP2}.  We now put it together with the previous results,
and, introducing the notation $\e=c_1\log (m/(c_1\cm))$ for brevity,
we arrive at the following estimates
\begin{eqnarray}
  \frac{\d M_\p^2}{M_\p^2} &\sim& \frac{M_\p^2}{N_f(4\p F_\p)^2}
  \ \sim \ \frac{\e}{N_f N_c} \ \sim \ \frac{\e}{N^2} \ ,
\label{dMpi}\\
  \frac{\d F_\p}{F_\p} &\sim& \frac{N_f M_\p^2}{(4\p F_\p)^2}
  \ \sim \ \frac{N_f}{N_c}\,\e \ \sim \ \e \ ,
\label{dFpi}\\
  \frac{\d M_\t^2}{M_\t^2} &\sim& \frac{N_f^2 M_\p^2}{(4\p F_\t)^2}
  \ \sim \ \frac{N_f^2}{N_c^2}\,\e \ \sim \ \e \ .
\label{dMtau}
\end{eqnarray}
This confirms that the one-loop corrections for these quantities
either stay finite or tend to zero in the Veneziano limit.\footnote{
  The presence of an extra $1/N_f$ suppression in the second
  expression in Eq.~(\ref{dMpi}) appears to be a peculiarity of the one-loop
  expression for $\d M_\p^2/M_\p^2$ .}
We expect that higher-order loop corrections will exhibit a similar behavior.
For the concrete case of $N_c=3$ and $N_f=8$ the ratio $N_f/N_c$
is quite large, consistent with the observation of Ref.~\cite{AIP2}
that the one-loop corrections for $\d F_\p/F_\p$ and $\d M_\t^2/M_\t^2$
could be relatively large.

%%%%%%%%%%%%%%%%%%%%%%%%%%%
%\newpage
\section{\label{LSD} Analysis of $N_f=8$ data}
%%%%%%%%%%%%%%%%%%%%%%%%%%%
In order to test the physical picture presented in the previous section,
we use numerical results obtained for the $N_f=8$, SU(3) gauge theory
in Ref.~\cite{LSD},\footnote{
  An update of these results has appeared recently \cite{LSD2}.
  This does not affect the estimates we make in this paper.
}
which used staggered fermions at a single value
of the bare coupling $\b$.  The applicability of the dilaton-pion
effective theory to these data was previously considered
in Ref.~\cite{AIP1,AIP2}.

The hyperscaling relations discussed in the previous section
imply that the ratio $M/F_\p$ should be roughly independent of $m$
for every hadron mass $M$ ($M_\p$ and $M_\t$ included).
The approximate constancy of this ratio for different hadrons
is evident from Fig.~4 of Ref.~\cite{LSD}.

%%%%%%%%%%%%%%%%%%%%%%%%%%%%%%%%%%%%%%%%%%%%%%%%%%%%%%%%%%%%%%%%%%%%%%%
\begin{table}[t]
\vspace*{2ex}
\begin{center}
\begin{tabular}{|c|c|c|c|c|r|}
\hline
$i$ & $am$ & $aM_N$ & $v(m_i)$ & $v(m_1)$ & $e^{2v(m_i)}$ \\
\hline
1 & 0.00125 & 0.25 & 2.0 &     &  57 \\
2 & 0.00223 & 0.32 & 2.4 & 2.3 & 120 \\
3 & 0.00500 & 0.44 & 2.5 & 1.9 & 150 \\
4 & 0.00750 & 0.52 & 2.6 & 1.9 & 180 \\
5 & 0.00889 & 0.58 & 3.4 & 2.6 & 900 \\
\hline
\end{tabular}
\end{center}
\begin{quotation}
\floatcaption{tabv}{$v(m)$ as a function of $m$, using results
for the nucleon mass from Ref.~\cite{LSD}, and assuming $y=2$.
The 5th column gives the value of $v(m_1)$ if the pair $(m_1,m_i)$ is used
to solve Eqs.~(\ref{Nmass}) and~(\ref{LOvev}).
For $v(m_1)$ in the 4th column we took the average of the values
obtained from the pairs $(m_1,m_2)$, $(m_1,m_3)$ and $(m_1,m_4)$.
Except for the second column, we keep only two significant digits.
Statistical errors have been suppressed.}
\end{quotation}
\vspace*{-3ex}
\end{table}
%%%%%%%%%%%%%%%%%%%%%%%%%%%%%%%%%%%%%%%%%%%%%%%%%%%%%%%%%%%%%%%%%%%%%%%

In order to be more quantitative, we have used the value $y=2$
(or, equivalently, $\g_*=1$) extracted in Ref.~\cite{AIP2}.  Given this value,
we can estimate the values of $v(m)$ for a given pair of masses
$m_i$ and $m_j$, using the values of the nucleon mass at the same
input masses reported in Fig.~1 of Ref.~\cite{LSD},
with the help of Eqs.~(\ref{LOvev}) and~(\ref{Nmass}).
The results are shown in Table~\ref{tabv}.  It should be noted that
the determination of $v$ is very sensitive to the input value of $y$.
Changing $y$ by as little as 3\% can change the values of $v$
by up to 15\%.  Moreover, such a variation gets exponentially
magnified.  For example,
a downward change of 15\% in the value of $v(m_5)$
would reduce $e^{2v(m_5)}$ by roughly a factor 3.

Next we estimated the hatted low-energy parameters,
employing the estimated values for $v(m_i)$ shown in the 4th column of Table~\ref{tabv},
again using $y=2$.  In Table~\ref{tablargem},
$a\hB_\p$ and $c_1 a\hB_\t$ were computed using Eq.~(\ref{mintreepion})
and~(\ref{mintreetau}), respectively;  $c_1a\cm$ was computed using Eq.~(\ref{LOvev}),
and $a\hf_\p$ was computed using Eq.~(\ref{defF}).  Finally, $a\hf_\t$
was obtained by combining the previous results with Eq.~(\ref{calM}).

In theory, all the hatted parameters, as well as $c_1 a\cm$,
should be roughly independent of the input mass, because the leading-order
dependence of the corresponding physical parameters on $e^{v(m)}$, and thus $m$, has been removed.  Of all the quantities
shown in Table~\ref{tablargem}, the most stable one is $a\hB_\p$.
Other quantities show a varying degree of sensitivity to the input mass
used for their calculation.  We believe that this sensitivity is
in part due to the large range of $e^{2v(m)}$ values shown in the rightmost column
of Table~\ref{tabv}, which, in turn, is very sensitive to the determination
of $y$, as we have already mentioned.  We comment that while the data
of Ref.~\cite{LSD} for the scalar-meson mass have relatively large
statistical errors (they can be as large as 15\% -- 20\%),
this is not enough to explain the systematic differences in our
estimates of $c_1 a\hB_\t$ for different values of $m$.
It could be that this variation is in part due to next-to-leading
order effects, which, as discussed in the previous section,
could be particularly large for the mass of the dilatonic meson.

We can now also estimate when the theory will enter
the small-mass regime of the effective theory,
$m/(c_1\cm) \ll 1$.
If we use the lightest fermion mass to estimate this ratio
from Table~\ref{tablargem}, we obtain
\begin{equation}
\label{deep}
  \frac{m}{c_1\cm} \ = \ O(100) \ ,
\end{equation}
which suggests that the fermion mass $m$ would have to be smaller by two
orders of magnitude before we reach the small-mass regime
at the same bare coupling.

The results shown in Table~\ref{tablargem} suggest the ratio
$\hB_\p/\hf_\p$ is of order $10^3$, which is much larger than in QCD.
Since $\hB_\p = -\hat\S_0 / \hf_\p^2$, where $\hat\S_0$ is the condensate
per flavor in the chiral limit, this can be considered as a signal
of condensate enhancement.

%%%%%%%%%%%%%%%%%%%%%%%%%%%%%%%%%%%%%%%%%%%%%%%%%%%%%%%%%%%%%%%%%%%%%%%
\begin{table}[t]
\vspace*{2ex}
\begin{center}
\begin{tabular}{|c|c|c|c|c|c|c|c|}
\hline
i & $aM_\p$ & $aM_\t$ & $10^2\,a\hf_\p$ & $10^2\,a\hf_\t$ &
$a\hB_\p$ & $10^6\,c_1a^2\hB_\t$ & $10^6\,c_1a\cm$ \\
\hline
1 & 0.082 &      & 0.29 &      & 2.7 &     & 5.3 \\
2 & 0.11  & 0.12 & 0.25 & 0.96 & 2.7 & 5.5 & 3.8 \\
3 & 0.17  & 0.21 & 0.32 & 1.1  & 2.8 & 12  & 6.6 \\
4 & 0.20  & 0.28 & 0.35 & 1.1  & 2.8 & 17  & 7.4 \\
5 & 0.23  & 0.24 & 0.17 & 0.71 & 2.9 & 1.9 & 1.4 \\
\hline
\end{tabular}
\end{center}
\begin{quotation}
\floatcaption{tablargem}{Values of the hatted low-energy constants,
assuming $y=2$.  We compute $a\hf_\p$ using Eq.~(\ref{defF}),
with $v(m_i)$ from the 4th column of Table~\ref{tabv},
while $a\hB_\p$ and $c_1 a\hB_\t$ are computed
using Eq.~(\ref{mintreepion}) and Eq.~(\ref{mintreetau}), respectively.
$c_1a\cm$ is computed using Eq.~(\ref{LOvev}).
$a\hf_\t$ in the 5th column is computed from columns 4,6,7, and 8,
using Eq.~(\ref{calM}).}
\end{quotation}
\vspace*{-3ex}
\end{table}
%%%%%%%%%%%%%%%%%%%%%%%%%%%%%%%%%%%%%%%%%%%%%%%%%%%%%%%%%%%%%%%%%%%%%%%

%%%%%%%%%%%%%%%%%%%%%%%%%%%
%\newpage
\section{\label{conc} Discussion and conclusion}
%%%%%%%%%%%%%%%%%%%%%%%%%%%

We finally return to a question that we have postponed until now,
which is how to identify the confinement scale of the massless theory.
Considering first the case of QCD,
both the pion decay constant and the strange quark mass
are of order 100~MeV, and moreover, it is well known that around
the strange mass higher-order chiral corrections become large,
and chiral perturbation theory may start to break down.
We thus propose to identify $\L$, the characteristic scale
of the massless theory, with $\hf_\p/N_c^{1/2}$.
If we were dealing with a QCD-like theory, the chiral expansion would
then start to break down for $mN_c^{1/2}/\hf_\p\sim 1$.

In a nearly conformal, but confining, theory
we may use the fact that $c_1 \sim n_f-n_f^*$ is parametrically small
to identify the following regions:
\begin{eqnarray}
  &\mbox{region A:}& \quad 0 \le m \ \ll\ c_1\cm \ ,
\label{regions}\\
  &\mbox{region B:}& \quad m \sim c_1\cm \ ,
\nonumber\\
  &\mbox{region C:}& \quad c_1\cm \ \ll\ m \ \ll\ \hf_\p/N_c^{1/2} \ ,
\nonumber\\
  &\mbox{region D:}& \quad m \sim \hf_\p/N_c^{1/2} \ ,
\nonumber\\
  &\mbox{region E:}& \quad \hf_\p/N_c^{1/2} \ \ll\ m \ \ll\ c_1\cm e^{1/c_1} \ .
\nonumber
\end{eqnarray}
In a QCD-like theory the scale $c_1\cm$ is not relevant,
and chiral perturbation theory is valid in the union of regions A, B, and C.
By contrast, in a near-conformal and confining theory,
the scale $\hf_\p$ does not play any special role.
This is a surprising result, because, based on QCD experience,
we had assumed in Ref.~\cite{PP} that $m/\L\equiv mN_c^{1/2}/\hf_\p$ must be small
in order for the effective theory to apply.

What happens instead,
is that the only relevant separation is between the small-mass regime,
region A, and the large-mass regime, which corresponds to the union of regions
C, D, and E,\footnote{
  Incidentally, the 8-flavor SU(3) gauge theory
  has $m/\hf_\p \sim O(1)$ \cite{LSD}.
}
as follows from conditions~(\ref{largemass}) and~(\ref{c1logm}).
In the small-mass regime, the dilatonic meson
decouples from the pions,
and the familiar chiral behavior is recovered.
Next comes the intermediate range $m\sim c_1\cm$.
Here the dilatonic meson might be as light as the pions,
but hyperscaling relations have not set in yet.  Beyond that we find
the large-mass regime, in which hadron masses and decay constants
depend to leading order on the fermion mass $m$
through $m^{\frac{1}{1+\g_*}}$.  This is recognized as the familiar
hyperscaling relation of a mass-deformed conformal theory
(see, for example, Ref.~\cite{luigi}),
except that here it occurs in a confining and chirally broken theory.
The pions and the dilatonic meson satisfy the same hyperscaling relation,
but they remain special in being much lighter than all other states.
In the large-mass regime, the familiar chiral expansion
in powers of $m$ is replaced by an expansion
in powers of $c_1\log(m/(c_1\cm))$.  The upper bound of the
large-mass regime stems from the logarithmic dependence on $m$ of the
new expansion parameter.  This logarithmic, instead of linear,
dependence on $m$ is what allows the large-mass regime
to extend over many scales.

The 8-flavor SU(3) gauge theory is qualitatively different from QCD
in a number of ways.  First, it contains a scalar meson which is
about as light as the pions in the range of fermion masses currently probed by
numerical simulations.  Second, in this mass range,
its spectrum satisfies approximate hyperscaling relations
which are characteristic of a mass-deformed conformal theory.

In this paper we considered these results using the low-energy effective theory
for pions and a dilatonic meson developed in Ref.~\cite{PP}.
We found that, even if the fermion mass is not small relative to
the characteristic infrared scale $\L$ of the massless theory,
the effective theory can still provide a systematic expansion
thanks to the parametric proximity of the conformal window,
quantified by the smallness of the expansion parameter $n_f-n_f^*$.
We found rough agreement between
the numerical results of Ref.~\cite{LSD} and the predictions of
the low-energy theory in this large-mass regime.

To date, there is fairly general consensus that the massless
8-flavor SU(3) gauge theory is confining.  However, the alternative scenario,
that this theory is infrared conformal, has not been ruled out.
The existence of a mass range where infrared conformal and confining
theories both exhibit similar hyperscaling relations to leading order
provides a possible explanation of why it can be so difficult to distinguish between the two scenarios numerically.
In also means that,
with currently accessible values of the fermion mass, decisive conclusions
cannot be reached unless subleading effects will be incorporated
in the analysis.
Under the hypothesis of a mass-deformed infrared conformal theory,
this amounts to the inclusion of (marginally) irrelevant operators
in the scaling analysis (see, for example, Ref.~\cite{Annaetal}).
The alternative hypothesis amounts to being in the large-mass regime
of the dilaton-pion effective theory, which is the subject of this paper.
Here, subleading effects include corrections to the approximate
classical solution~(\ref{largemsol}), which are discussed in App.~\ref{v012},
as well as the usual next-to-leading and higher order loop corrections.
Of course, besides the subleading effects of the continuum
low-energy theory, discretization effects should be taken into account as well.

One could in principle show that the 8-flavor SU(3) gauge theory
is confining by reaching the small-mass region of the effective theory
(region A of Eq.~(\ref{regions})), where
the pion mass exhibits its usual chiral behavior,
while the masses of all other hadrons, including the dilatonic meson,
are non-zero, and independent of the fermion mass to leading order.
However, we estimated that in order to reach this regime at the
value of the bare coupling used in Ref.~\cite{LSD}, the fermion mass
would have to be smaller at least by two orders of magnitude.

The SU(3) gauge theory with $N_f=2$ Dirac fermions in the sextet representation
is, strictly speaking, not within the scope of the effective theory,
since a Veneziano limit cannot be taken for matter fields in
two-index representations.  Nevertheless, we speculated in Ref.~\cite{PP}
that the effective theory might be applicable under the assumption
that $N_f-N_f^*$ is small, where $N_f^*$ is the (non-integer) value
where the (non-local) SU(3) gauge theory with $N_f^*$ sextet fermions
enters the conformal window.  The predictions of (this version
of) the effective theory were compared with numerical data in
Ref.~\cite{sextet}, and the emerging physical picture,
including both the general overall agreement,
and the need to sort out many issues in more detail, is rather similar
to the case of $N_f=8$ fundamental flavors.

%%%%%%%%%%%%%%%%%%%%%%%%%%%
\vspace{2ex}
%\newpage
\noindent {\bf Acknowledgments}
\vspace{2ex}
%%%%%%%%%%%%%%%%%%%%%%%%%%%

This material is based upon work supported by the U.S. Department of
Energy, Office of Science, Office of High Energy Physics, under Award
Number DE-FG03-92ER40711 (MG).   MG
would like to thank the Instituto de Fisica Teorica (IFT UAM-CSIC) in Madrid
for its support via the Centro de Excelencia Severo Ochoa Program
under Grant SEV-2016-0597.
YS is supported by the Israel Science Foundation
under grant no.~491/17.

%%%%%%%%%%%%%%%%%%%%%%%%%%%
%\newpage
\appendix
\section{\label{classical} Revisiting the classical solution}
%%%%%%%%%%%%%%%%%%%%%%%%%%%
Before we use the freedom to shift the $\t$ field,
the leading-order lagrangian is
\begin{equation}
  \tcl = \tcl_\p + \tcl_\t + \tcl_m + \tcl_d \ ,
\label{LeffL}
\end{equation}
where
\begin{eqnarray}
  \tcl_\p &=& \frac{f_\p^2}{4}\, V_\p(\t)\, e^{2\t}
              \tr(\partial_\m \S^\dagger \partial_\m \S) \ ,
\label{LpV}\\
  \tcl_\t &=&
  \frac{f_\t^2}{2}\, V_\t(\t)\, e^{2\t} (\partial_\m \t)^2  \ ,
\label{LtV}\\
  \tcl_m &=& -\frac{f_\p^2 B_\p}{2} \, V_M(\t)\, e^{y\t}
  \tr\Big(\c^\dagger \S + \S^\dagger \c\Big) \ ,
\label{LmV}\\
  \tcl_d &=& f_\t^2 B_\t \, e^{4\t} V_d(\t) \ .
\label{LdV}
\end{eqnarray}
Each potential $V_\p$, $V_\t$, $V_M$ and $V_d$ has a double expansion
in powers of $\t$ and of $n_f-n_f^*$.
According to the power-counting arguments of Ref.~\cite{PP}, the power
of $n_f-n_f^*$ cannot be smaller than the power of $\t$.  In particular,
\begin{equation}
\label{Vdnk}
  V_d = \sum_{n=0}^\infty \sum_{k=n}^\infty \tc_{nk} \t^n (n_f-n_f^*)^k \ .
\end{equation}
With the power counting in place, we still have the problem that
for a generic potential $V_d(\t)$ we
anticipate the expectation value of the dilatonic meson to behave as
\begin{equation}
\label{vOnf}
  \svev{\t} = O(1/(n_f-n_f^*)) \ .
\end{equation}
Therefore, {\em a-priori} we do not know what is the $O(1)$ part
of each potential.  In this appendix, we consider this issue in more
detail.

In the massless limit, the classical potential of the dilatonic meson is
\begin{subequations}
\label{Vdeff}
\begin{eqnarray}
  V_{cl}(\t) &=& f_\t^2 B_\t U(\t) \ ,
\label{Vdeffa}\\
  U(\t) &=& V_d(\t) e^{4\t} \ .
\label{Vdeffb}
\end{eqnarray}
\end{subequations}
We begin by reorganizing the expansion of $V_d$ as
\begin{eqnarray}
\label{Vdx}
  V_d &=& \sum_{m=0}^\infty (n_f-n_f^*)^m V_m(x) \ ,
\\
  V_m(x) &=& \sum_{n=0}^\infty \tc_{n,n+m} (n_f-n_f^*)^n \t^n
  = \sum_{n=0}^\infty \tc_{n,n+m} x^n \ ,
\nonumber
\end{eqnarray}
where we have introduced the variable
\begin{equation}
\label{defx}
  x = (n_f-n_f^*) \t \ .
\end{equation}
When we are dealing with the classical solution, for which $\t=O(1/(n_f-n_f^*))$,
we have that $x=O(1)$, and thus Eq.~(\ref{Vdx}) gives the relevant
expansion of $V_d$ in powers of $n_f-n_f^*$.
Each $V_m(x)$ is then some $O(1)$ function of its argument.

Following Ref.~\cite{Rattazzi}, we now make the assumption
that there exists an $x_0$ such that
\begin{equation}
\label{V0x}
  V_0(x_0)=0 \ .
\end{equation}
For simplicity, we make some further technical assumptions as well.
First, we assume that, in Eq.~(\ref{Vdeffb}), the exponential $e^{4\t}$
dominates over $V_d(\t)$ for $\t\to\pm\infty$.  This implies
that $U(\t)\to 0$ for $\t\to -\infty$.   In addition, we assume
that $V_d(\t)$ is positive for $\t\to\infty$, so that $U(\t)\to+\infty$
for $\t\to+\infty$.  Finally, we assume that the zero of $V_0$ is unique,
which in turn implies that the unique saddle point found below
is the global minimum of $U(\t)$.

We will now demonstrate the existence of a stable classical solution.
To this end, we shift the dilatonic meson field as
\begin{equation}
\label{shift}
  \t \to \t + \D \ , \qquad \D =  \frac{x_0}{n_f-n_f^*} \ .
\end{equation}
The shift entails several rearrangements in the tree-level lagrangian.
The low-energy constants are redefined according to $f_{\p,\t}\to e^\D f_{\p,\t}$,
$B_\p\to e^{(y-2)\D}B_\p$, and $B_\t\to e^{2\D}B_\t$.  Notice that these
redefinitions depend on $n_f$, and thus they have to be taken into account
when comparing theories with different $n_f$.  The $\t$ shift also gives rise
to a rearrangement of the expansions of the potentials.
In terms of the $x$ variable, the shift takes the simple form $x\to x+x_0$.
Considering the expansion of $V_d$ in Eq.~(\ref{Vdx}), in effect the shift
implies that we are now expanding around $x=0$, instead of around $x=x_0$
(a similar statement applies to the other potentials).  Explicitly, we have
\begin{equation}
\label{Vxx0}
  V_m(x) \equiv V_m^{\rm orig}(x+x_0) \equiv \sum_{n=0}^\infty c_{mn} x^n \ ,
\end{equation}
where $V_m^{\rm orig}$ denotes the original form of $V_m(x)$ before the shift.

From now on we assume that the shift~(\ref{shift}) has been carried out.
Equation~(\ref{V0x}) thus takes the simple form
\begin{equation}
\label{V0shift}
  V_0(0)=0 \ .
\end{equation}
In terms of the new expansion coefficients introduced in Eq.~(\ref{Vxx0}),
this implies $c_{00}=0$.

Let us now find the classical solution $\tx$.
As we will see, the classical solution satisfies $\tx=O(n_f-n_f^*)$.
Moreover, the classical solution is stable,
namely, it admits an expansion in powers of $n_f-n_f^*$.
Assuming self-consistently that $\tx=O(n_f-n_f^*)$ we have,
neglecting terms of $O((n_f-n_f^*)^2)$ as we go,
\begin{eqnarray}
\label{Uprime}
  U'(\t) &=& [4V_0(x) + (n_f-n_f^*)(V'_0(x)+4V_1(x))] e^{4\t}
\\
  &=& [4c_{01}x + (n_f-n_f^*)(c_{01}+4c_{10})] e^{4\t} \ ,\nonumber
\end{eqnarray}
where a prime denotes differentiation with respect to the argument.
To this order the classical solution is therefore
\begin{equation}
\label{xeps}
  \tx = (n_f-n_f^*) \tx_1 \ , \qquad
  \tx_1 = - \frac{c_{01}+4c_{10}}{4c_{01}} \ .
\end{equation}
The generalization to higher orders is straightforward.

The shift has removed the large, $O(1/(n_f-n_f^*))$ component
of the $\t$ field.  The leading-order classical solution after the shift
is $\svev\t=\tx_1$, which is $O(1)$.
Likewise, the quantum field is $O(1)$, as always in perturbation theory.
Thus, no inverse powers of $n_f-n_f^*$ are hidden in the $\t$ field any more,
and we may truncate each potential according to the explicit power of
$n_f-n_f^*$ which should be kept at a given order of the low-energy expansion.
In particular, we may truncate $V_\p$, $V_\t$ and $V_M$ to their leading-order
constant value, which in turn we normalize to unity.

As for $V_d$, it is convenient to do a second, $O(1)$ shift of the $\t$ field.
After the shift~(\ref{shift}), $U(\t)$ is given at order $n_f-n_f^*$ by
\begin{equation}
\label{Uepstau}
  U(\t) = (n_f-n_f^*) (c_{01}\t + c_{10}) e^{4\t} \ .
\end{equation}
The second shift is $\t\to\t+\tD$ with
\begin{equation}
\label{shift2}
  \tD = -\frac{1}{4}-\frac{c_{10}}{c_{01}} \ .
\end{equation}
This gives rise to
\begin{equation}
\label{newU}
  U(\t) = (n_f-n_f^*) c_{01} (\t-1/4) e^{4(\t+\tD)} \ ,
\end{equation}
and so at this order the classical solution is $\svev{\t}=0$,
equivalently $\tx_1=0$.  The resulting tree-level lagrangian
is given by Eqs.~(\ref{Leff}) through~(\ref{Ld}), where in Eq.~(\ref{Ld})
$c_1=(n_f-n_f^*) c_{01}$.

The analysis in this appendix has so far been carried out
while invoking the Veneziano limit, $N\to\infty$.
This analysis carries over to $0<1/N\ll 1$, provided we make a further,
technically reasonable assumption that the only inverse small parameter
occurring in the classical solution is $1/(n_f-n_f^*)$, see Eq.~(\ref{vOnf}).
Instead of Eq.~(\ref{Vdx}), $V_d$ now admits the expansion
\begin{equation}
\label{VdN}
  V_d= \sum_{m,k=0}^\infty (n_f-n_f^*)^m N^{-k}\, V_{mk}(x) \ ,
\end{equation}
where the $V_{mk}(x)$ are $O(1)$ functions of their argument,
and, after performing the large shift of Eq.~(\ref{shift}), we have $V_{00}(0)=0$,
generalizing Eq.~(\ref{V0shift}).
It is then straightforward to check that the classical solution remains stable,
and admits a double expansion in powers of both $(n_f-n_f^*)$ and $1/N$.

%%%%%%%%%%%%%%%%%%%%%%%%%%%
%\newpage
\section{\label{v012} Large-mass expansion of the classical solution}
%%%%%%%%%%%%%%%%%%%%%%%%%%%
In this appendix we work out corrections to the approximate
classical solution $v_0(m)$ in the large-mass regime.
We show that the exact classical solution $v(m)$ admits an expansion
in inverse powers of $v_0 \sim \log m$, with coefficients that
are polynomials in $\log v_0 \sim \log\log m$.

We start from the solution of the approximate equation~(\ref{largemsol}),
namely,
\begin{equation}
\label{v0}
  v_0 = \frac{1}{1+\g_*}\log \frac{ym}{4c_1\cm} \ .
\end{equation}
Writing $v=v_0+v_1+\cdots$, and substituting this into
the exact saddle-point equation~(\ref{LOvev}),
the next correction $v_1$ satisfies
\begin{equation}
\label{NLOvev}
  e^{(1+\g_*)v_0} = e^{(1+\g_*)(v_0+v_1)+\log v_0}\ ,
\end{equation}
hence
\begin{equation}
\label{v1vev}
  v_1 = -\frac{\log v_0}{1+\g_*} \ .
\end{equation}
It follows that
\begin{equation}
\label{v1}
  e^{v_0+v_1} = e^{v_0}\, v_0^{-(1+\g_*)} \ .
\end{equation}
Note that $v_1$ does not tend to zero for $m\to\infty$,
but, as expected, $v_1/v_0 \sim \log v_0/v_0$.
Proceeding to the next correction and using Eq.~(\ref{v1vev}) gives
\begin{equation}
\label{v2vev}
  v_2 = %-\frac{v_1}{(1+\g_*)v_0}
  \frac{\log v_0}{(1+\g_*)^2\,v_0} \ .
\end{equation}
The last correction we calculate is
\begin{equation}
\label{v3vev}
  v_3 = \frac{\log^2 v_0 - 2\log v_0}{2(1+\g_*)^3\,v_0^2} \ .
\end{equation}
The corrections $v_2,v_3,\ldots,$ tend to zero for $v_0\to\infty$.

%\newpage
\vspace{3ex}
%%%%%%%%%%%%%%%%%%%%%%%%%%%

\end{document}